\documentclass{Interspeech2024}

\usepackage{soul}

\interspeechcameraready

\title{Infusing Acoustic Pause Context into Text-Based Dementia Assessment}
\name[affiliation={1}]{Franziska}{Braun}
\name[affiliation={2}]{Sebastian P.}{Bayerl}
\name[affiliation={4}]{Florian}{Hönig}
\name[affiliation={3}]{Hartmut}{Lehfeld}
\name[affiliation={3}]{Thomas}{Hillemacher}
\name[affiliation={1}]{Tobias}{Bocklet}
\name[affiliation={1}]{Korbinian}{Riedhammer}

\address{
  $^1$Technische Hochschule Nürnberg,
  $^2$Technische Hochschule Rosenheim, Germany\\
  $^3$Klinik für Psychiatrie und Psychotherapie, Universitätsklinik der Paracelsus Medizinischen Privatuniversität, Klinikum Nürnberg, Germany, 
  $^4$KST Institut GmbH, Bad Emstal, Germany}
\email{franziska.braun@th-nuernberg.de}

\begin{document}

\maketitle

\begin{abstract}
Speech pauses, alongside content and structure, offer a valuable and non-invasive biomarker for detecting dementia. 
This work investigates the use of pause-enriched transcripts in transformer-based language models to differentiate the cognitive states of subjects with no cognitive impairment, mild cognitive impairment, and Alzheimer's dementia based on their speech from a clinical assessment. 
We address three binary classification tasks: Onset, monitoring, and dementia exclusion. 
The performance is evaluated through experiments on a German Verbal Fluency Test and a Picture Description Test, comparing the model's effectiveness across different speech production contexts. 
Starting from a textual baseline, we investigate the effect of incorporation of pause information and acoustic context.
We show the test should be chosen depending on the task, and similarly, lexical pause information and acoustic cross-attention contribute differently.
\end{abstract}

\section{Introduction}
%Motivation 
Recent breakthroughs in antibody therapy offer promising prospects for slowing the progression of Alzheimer's disease when applied at an early stage \cite{van_dyck_lecanemab_2023}. 
However, early and accurate detection of dementia and monitoring of disease progression remains a major challenge in clinical assessment. 
Current gold standard biomarkers, such as those from blood, cerebrospinal fluid and magnetic resonance imaging (MRI), offer the most accurate detection of Alzheimer's disease, but are invasive (except MRI), expensive and often not easily accessible.
Furthermore, as they are not sensitive enough for early stages of dementia such as mild cognitive impairment, a combination of clinical assessment and neuropsychological evaluation remains crucial for accurate and early diagnosis.
Speech analysis has proven to be a promising non-invasive and cost-effective approach to detecting dementia.
It also has the advantage of providing biomarkers that are easily accessible during neuropsychological assessment or through spontaneous speech.

In this paper, we use speech samples from a German multi-center study in which established clinical assessment tools were used to assess the speech of individuals with no cognitive impairment (NC), with mild cognitive impairment (MCI), and with mild to moderate dementia (AD) in the context of Alzheimer's disease.
To examine the performance of the models in different contexts of speech production, we compare the speech from two established cognitive tests: a Verbal Fluency Test (VFT) and a Picture Description Test (PDT).
Our experiments aim to distinguish speech markers of individuals in onset dementia (NC vs. MCI), in the monitoring of dementia (MCI vs. AD) and in the exclusion of dementia versus healthy controls (NC vs. AD).
%with NC from those with MCI (early-stage dementia detection)
%with MCI from those with AD (dementia progression monitoring)
%with NC from those with AD (dementia exclusion)

To achieve this, we extract speech markers that go beyond the content and structure of what is said and additionally look at what is not said: the pauses in speech.
Speech pauses can be a valuable indicator of dementia as they differ in frequency, duration and other characteristics (e.g., syntax position) for different stages of cognitive decline.
We investigate different approaches to encode pauses in language models and in a first approach we train a self-attention based system to learn the pause context from the text modality.
In a second approach, we allow the text system to incorporate the pause context from acoustic information using cross-attention.

In this paper, we analyze different ways to incorporate pause duration and acoustic context.
We find that:
\begin{enumerate}
    \item NC and MCI are best discriminated using the VFT when acoustic information is incorporated. % outlook: masking to benefit from pauses?
    \item MCI and AD can best be distinguished using the PDT, whereby the modeling of disfluencies and pauses in the text-based system is sufficient.
    \item NC and AD can be reliably distinguished independently of the test, but the modeling of pauses is beneficial.
\end{enumerate}

\section{Related Work}
%Speech Pauses in Dementia Detection
Speech analysis has shown promise as a non-invasive approach to detecting dementia, with an increasing focus on investigating the potential of pauses in speech as a valuable indicator.

Several studies have observed a correlation between the severity of dementia and increased pause duration in various speech tasks, including storytelling \cite{vincze_telltale_2021}, picture description \cite{sluis_automated_2020, pastoriza-dominguez_speech_2022, yuan_disfluencies_2020, yuan_pause-encoded_2021}, verbal fluency \cite{lofgren_breaking_2022, koenig18}, reading \cite{meilan_speech_2014}, recall tasks \cite{vincze_telltale_2021} and many more. 
This suggests that analyzing pause patterns could provide insights into cognitive function and potentially help distinguish healthy individuals from those with dementia.

Beyond the overall duration, researchers are investigating the specific characteristics of pauses at different stages of Alzheimer's dementia (AD). 
For example, some studies suggest that people with AD show longer silent pauses (related to word-finding problems) than healthy controls.
While Lofgren et al. \cite{lofgren_breaking_2022} found no significant difference in the frequency or duration of filled pauses, Yuan et al. \cite{yuan_disfluencies_2020} found that individuals with AD tended to use the filler ``uh'' more frequently than healthy individuals.
Syntax-related analyses showed that certain verbs (such as ``is'') are sensitive to AD \cite{zhu_towards_2022} or that clause-initial pauses occur more frequently in moderate AD, while more pauses occur within-clause positions in mild AD, especially before nouns (challenges in referencing objects) \cite{lofgren_breaking_2022}.

In many studies, features from speech analysis (e.g., lexical, linguistic) and signal processing (e.g., acoustic, temporal) were combined to classify NC, MCI and AD, often including pause statistics such as pause rate and duration \cite{koenig18, konig_use_2018, calza_linguistic_2021, asgari_predicting_2017, nagumo_automatic_2020, vincze_linguistic_2022}.
König et al. \cite{koenig15} found that in picture description tasks, features that reflect the continuity of speech (i.e., longer contiguous speech segments and shorter silence segments) have the highest discriminative power.
For semantic fluency, the largest contribution to classification accuracy was obtained from the temporal positions of individual words in the first part of the task. 
Features obtained from the duration of speech and silence segments were useful for discriminating MCI from AD, but were not significant for NC vs. MCI.

Other studies showed the suitability of transformer-based architectures that encode linguistic (e.g., BERT) and acoustic (e.g., W2V2) contextual information for the detection of cognitive impairment \cite{braun_GoingCookieTheft_2022, braun_classifying_2023, balagopalan_bert_2020}.
Yuan et al. \cite{yuan_disfluencies_2020, yuan_pause-encoded_2021} presented a method for encoding filled and unfilled pauses in transcripts of picture descriptions to fine-tune the training of language models (ERNIE, BERT and RoBERTa), and achieved high accuracies in the detection of AD, for example in the ADReSS challenge \cite{yuan_disfluencies_2020, adress20}.
In our previous work \cite{braun_classifying_2023}, we found that although we achieved high performance in dementia classification using BERT and W2V2, distinguishing MCI from higher and lower cognitive impairment classes remained challenging and may be related to other neurodegenerative disorders (e.g., depression).

Most recent work in the TAUKADIAL challenge of INTERSPEECH 2024 introduced a multilingual baseline for MCI detection using speech from a novel benchmark of picture description tasks in English and Chinese.
The authors used established baselines such as audio signal, linguistic and acoustic features and reported results that were around chance level. 
They obtained the best MCI classification results of 59.18\% (UAR) by combining eGeMAPs and W2V2 features \cite{LuzEtAlTAUKADIAL24}.
This highlights the ongoing major challenges associated with generalizable early-stage dementia detection.

In this work, we aim to fill the gap in literature by investigating pause-enhanced transformers for the detection of NC, MCI and AD.
These pause-enhanced transformers combine the aforementioned approaches from speech analysis and signal processing by inherently capturing linguistic, syntactic, acoustic and temporal features and additionally enriching them with pause information.

\section{Data}\label{sc:data}
The data was selected from a multi-center study that was presented and described in detail in our previous work \cite{braun_classifying_2023}; it was provided by the PARLO Institute for Research and Teaching in Speech Therapy.
This study is an open, controlled, cross-sectional clinical study with parallel groups, conducted with the same iPad type and app at nine academic memory clinics across Germany.
The multi-center study allows for generalizability across populations, dialects and recording conditions and has shown to generalize across other independently recorded German datasets \cite{braun_classifying_2023}.
The demographic data of the 82 NC, 58 MCI and 65 AD German-speaking subjects can be found in \tablename~\ref{tab:demo}.

\begin{table}[th]
  \caption{Demographics for no cognitive impairment (NC), mild cognitive impairment (MCI), Alzheimer's Dementia (AD) group}
  \label{tab:demo}
  \centering
    \begin{tabular}{c|ccc}
    \toprule
        & \textbf{Count}         & \textbf{Age}               & \textbf{Gender}   \\
    \midrule
    NC  & 82 & 55-87 (68.9+-7.9) & 32m/50f  \\
    MCI & 58 & 55-85 (70.9+-8.3) & 31m/27f  \\
    AD & 65 & 55-85 (71.6+-8.8) & 32m/33f  \\
    \midrule
    All & 205 & 55-87 (70.3+-8.3) & 95m/110f \\
    \bottomrule
    \end{tabular}
\end{table}

The speech samples used contain approximately one minute of audio per subject during the performance of a semantic Verbal Fluency Test (VFT) and one minute during the performance of a Picture Description Test (PDT).

The semantic VFT is often used in the diagnosis of various dementia sub-types, especially in the early stages.
It measures the speed and ease of verbal production, semantic memory, linguistic abilities, executive functions and cognitive flexibility \cite{isaacs73}.
The task of this VFT is to name as many terms as possible from a semantic category in a given time limit.
In this case, the subject was asked to name as many different animals as possible within one minute.

\begin{figure}[th]
  \centering
  \includegraphics[width=\linewidth]{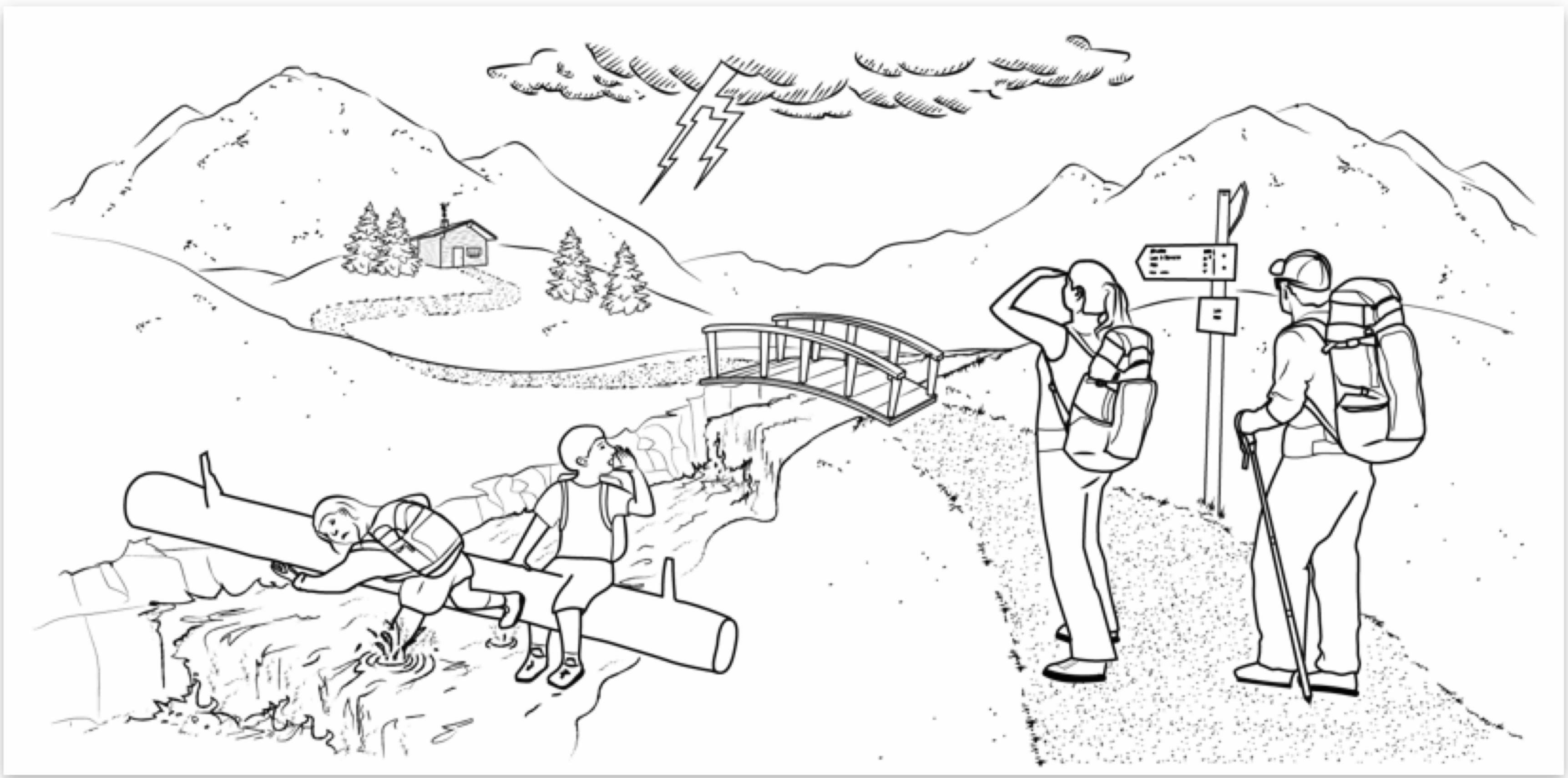}
  \caption{Picture Description Task ``Mountain Scene''}
  \label{fig:mountain}
\end{figure}

The PDT was developed primarily for the detection of Alzheimer's disease. 
It measures the amount and quality of information that a subject can obtain from a visual stimulus by describing a picture. 
In this case, the subject was asked to describe the picture of a mountain scene as shown in \figurename~\ref{fig:mountain}\footnote{PARLO GmbH, \url{www.parlo-institut.de}}.

% age mean 70.3+-8.3 min 55-87 NC: 68.9+-7.9 55-87, 31m/50f MCI: 70.9+-8.3 55-85, 31m/27f AD: 71.6+-8.8 55-85, 32m/33f

\section{Method}\label{sc:method}
\subsection{Pause-Enriched Transcripts}\label{sc:trans}
For our experiments, we create automatic transcripts using Whisper \cite{lintoai2023whispertimestamped}; accurate timestamps are computed using DTW applied to cross-attention weights, as demonstrated by \cite{JSSv031i07}.

For transcription, we use \texttt{whisper-large-v3} (beam\_size=5, best\_of=5, temperature=(0.0, 0.2, 0.4, 0.6, 0.8, 1.0)).
Prior to transcription, we run the embedded Silero VAD to prevent the whisper model from ``hallucinating'' text when a segment without speech is present and to obtain more accurate word timestamps.
Pause durations are computed across all segments between the end time of each word and the start time of the following word.
In order to add the resulting pauses to the transcripts, they were grouped by mapping them to the following pause duration intervals (in seconds):
% TABELLE?!
\begin{itemize}
    \item Baseline P1 (3 special tokens): $[0.05, 0.5[$; $[0.5, 2.0]$; $]2, \infty[$
    \item Baseline P2 (6 special tokens): $[0.05, 0.1]$; $]0.1, 0.3]$; $]0.3, 0.6]$; $]0.6, 1.0]$; $]1.0,2.0]$; $]2.0, \infty[$
    \item Baseline P3 (3 special tokens): $[0.2, 0.6]$; $]0.6, 1.5[$; $[1.5, \infty[$
    \item Baseline P4 (1 special token): $[0.2, 0.6]$; $]0.6, 1.5[$; $[1.5, \infty[$ 
\end{itemize}
The pause intervals in baseline P1 \cite{yuan_disfluencies_2020} were empirically derived from the data analysis of the ADReSS dataset \cite{adress20}, which is a subset of 156 speakers (NC and AD) of the Pitt corpus \cite{becker1994dementiabank} and contains recordings of the Cookie Theft Picture description task \cite{borod80}.
Baseline P2 \cite{yuan_pause-encoded_2021} proposed a pause distribution derived from the data-driven analysis of the INTERVIEW dataset \cite{majumder-etal-2020-interview}, which contains 105K conversations (10K hours) from seven National Public Radio programs over 20 years.
A linguistic study (baseline P3) \cite{pastoriza-dominguez_speech_2022} examined pauses in the speech of NC, MCI and AD individuals performing a narrative picture-story description task from the Bilingual Aphasia Test \cite{paradis1987bilingual}.
Thus, we obtain pause duration distributions for the groups of NC vs. AD (baseline P1), NC only (baseline P2) and NC vs. MCI vs. AD (baseline P3).
The resulting pauses are automatically inserted into the transcripts during the transcription process using special tokens.
Baseline P4 is obtained using the pause modeling of P3, but instead of adding three special tokens for each pause, we add only one special token at the respective positions regardless of duration. 
P4 therefore contains positional but no temporal pause information.
In addition, we use the library's implicit disfluency token ([*]), which is output when potential disfluencies in the attention weights are identified during speech recognition, and added it to P3 to obtain another baseline.
The reason we focus on P3 is that it is the only baseline that examines pauses in all three health conditions (HC, MCI and AD).

\subsection{Text Embeddings}\label{sc:bert}
We choose Bidirectional Encoder Representations from Transformers (BERT) as a pre-trained masked language model for natural language processing, as it has been shown to be suitable for our purposes in previous work  \cite{braun_classifying_2023, yuan_disfluencies_2020, yuan_pause-encoded_2021}.
BERT uses a transformer-based architecture to process large amounts of text data and capture the relationships between words in a sentence.
%BERT models can be adapted for a variety of NLP tasks, including question answering, sentiment analysis, and text classification. 
In our experiments, we use the BERT base model pre-trained on about 12 GB of German text data (Wiki, OpenLegalData, News) to predict masked words and the next sentence.
The model weights are open-source and can be accessed online \footnote{\url{https://huggingface.co/bert-base-german-cased}}.

Additional special tokens are added for all pause tokens from section~\ref{sc:trans} to the BERT Tokenizer and the model embeddings are adapted to the new vocabulary length.
We obtain the final text embeddings from the last hidden states of the model without pooling but with zero padding along the sequence dimension, which corresponds to 768 dimensional feature vectors per token in the input sequence. 

\subsection{Audio Embeddings}
The audio embeddings are derived from wav2vec 2.0 (W2V2), a transformer-based architecture designed for learning speech representations from raw audio data.
The model's convolutional layers act as a powerful feature encoder that processes raw audio waveforms directly. 
This allows the model to learn data-driven representations that can capture specific contextualized speech features obtained from 12 transformer blocks (W2V2-base) that use self-attention to focus on the task-specific parts of the audio.
The features extracted from W2V2 have been successfully used to detect cognitive impairment in previous work \cite{braun_classifying_2023, braun_GoingCookieTheft_2022, balagopalan_comparing_2021}.
We use a base model that was fine-tuned based on the Mozilla Foundation's Common Voice 9.0 dataset as a feature extractor without adjustment; the model weights are open-source and can be obtained online \footnote{\url{https://huggingface.co/oliverguhr/wav2vec2-base-german-cv9}}.

We z-normalized the waveform data before input and obtained a 768-dimensional speech representation after each transformer layer, representing about 0.02 seconds of the audio.
This results in $N = T/0.02 - 1$ vectors for the extraction context of $T$ (i.e., 449 vectors for an extraction context of 10 seconds).
To obtain the final embeddings, we compute the mean vector over all extracted feature vectors of a sample and apply zero-padding along the time dimension.

\section{Experiments}\label{sc:exp}
To assess the performance of the models, we use stratified five-fold cross-validation (5-fold CV) by splitting the data into five speaker-distinct training and test sets comprising ~80\% and ~20\% of the data, respectively.
We use the Area Under the Curve (AUC) to measure the ability of the binary classifiers to discriminate between classes as a summary of the ROC curve. 
The higher the AUC value, the better the performance of the model in discriminating between the positive and negative classes.

Our binary classification experiments aim to discriminate speech and speech pauses of individuals in three clinical assessment tasks: Dementia onset (NC vs. MCI), monitoring (MCI vs. AD) and exclusion (NC vs. AD).
All experiments are conducted with speech data from the two cognitive tests described in section~\ref{sc:data} using the text and audio embeddings from section~\ref{sc:method}.
BERT features are extracted using the pause-enriched transcripts from section~\ref{sc:trans} for P1--P4 as well as for P3 plus disfluency tokens.
For W2V2 features the extracted layer $L$ is selected from $L \in \{1, 2, \ldots, 12\}$.

Please note that for reasons of comparability, all experiments are performed in a fixed hyper-parameter setting (lr=5e-5, optimizer=Adam, batch\_size=8, max\_epochs=20, activation\_function=ReLU, drop\_out=0.1) without tuning.
We assume that optimizing the hyper-parameters can lead to higher performance.
The training is stopped early to prevent the model from possibly overfitting on the limited training data.
We use the cross-entropy loss function in which each class is assigned a rescaling weight to compensate with unbalanced class distributions during training.
The model architecture consists of a single-head attention module (pytorch), followed by a MLP classifier with one hidden layer (hidden\_dim=512) and one output layer with two output neurons.
Stratified batch sampling is used during training, where the batched data is first passed to the attention module.
Mean pooling and layer normalization are applied to the resulting attention outputs and then passed to the MLP classifier.
To calculate probabilities and class predictions softmax and argmax are applied to the logits of the model.

In our baseline experiments, we use BERT and W2V2 features separately and apply self-attention (i.e., query, key, and value remain the same) to learn text-based and acoustic-based pause context.
In an improved setting, we allow the text-based system to learn pause context from the acoustic embeddings by using one-way cross-attention from the text to the audio modality (i.e., using text as query and audio as key and value).

\section{Results}
%The binary classification results are shown in \tablename~\ref{tab:hcmci} NC vs. MCI, \tablename~\ref{tab:mciad} for MCI vs. AD and \tablename~\ref{tab:hcad} for NC vs. AD.
The binary classification results for the detection of early dementia are shown in \tablename~\ref{tab:hcmci}, with the best results obtained for VFT for the P4 baseline with the inclusion of acoustic context (cross-attention) and P3 modeling plus disfluencies in the text-based system (self-attention).
This could be related to the fact that P3 is the only one of the baselines that initially proposed pause distributions that are characteristic in the MCI condition. 
We hypothesize that the disfluencies in the text-based assessment together with the pause duration provide relevant information about verbal fluency, both of which could presumably also be learned from the acoustics in P4, the baseline without text-based pause duration information.
The fact that the VFT generally performs better than the PDT in this task could be due to the fact that it triggers speech production that is already impaired at early stages of dementia.

\tablename~\ref{tab:mciad} shows that MCI and AD are best distinguished using the PDT, where modeling of pauses (P3) and disfluencies in the text-based scenario is sufficient. 
As with early detection, we attribute this to the fact that P3 modeling relates to the MCI stage and that fluency is likely to decrease as cognitive decline progresses.
We suspect that the PDT works better than the VFT in this case because it is specialized for the detection of AD.

As shown in \tablename~\ref{tab:mciad}, NC and AD can be reliably distinguished regardless of the test, but modeling the pauses seems to be advantageous.
The best results for both tests, VFT and PDT, were obtained with with P1 modeling. 
We assume that this is due to the fact that the P1 baseline analyzed speech pauses in AD classification from healthy controls, so the task matches the investigated speech basis of the pause coding.

The W2V2 features performed worse overall than the best pause-enhanced rating in all tasks and tests, but often improved performance when used in cross-attention.

\begin{table}[th]
\caption{Average AUC (in \%) in 5-fold CV for the classification of \textbf{NC vs. MCI} using BERT and W2V2 features (best layer) from the VFT and PDT in self-attention and cross-attention with and without pause coding (P1--4) and disfluencies (Disfl.).}
  \label{tab:hcmci}
  \centering
\begin{tabular}{l||cc|cc}
            \toprule
            & \multicolumn{2}{c}{\textbf{VFT}}                                   & \multicolumn{2}{c}{\textbf{PDT}}                       \\
            \midrule
features      & self-att                        & cross-att                       & self-att                        & cross-att           \\
            \midrule
BERT             & 66.3                        & 71.0                        & 57.8                        & 67.5            \\
BERT P1          & 66.0                        & 70.2                        & 58.5                        & \textbf{70.3}             \\
BERT P2          & 65.5                        & \ul{71.2}                   & 52.4                        & \ul{70.0}  \\
BERT P3          & 66.7                        & \ul{71.2}                   & 56.6                        & 68.6  \\
BERT P4          & 66.1                        & \textbf{71.4}               & 57.1                        & 68.0  \\
BERT P3 + Disfl. & \textbf{69.3}               & 71.1                        & \ul{62.4}                   & 67.9  \\
            \midrule
W2V2             & \ul{67.5}                            & -                  & \textbf{66.2}                         & -           \\
          \bottomrule
\end{tabular}
\end{table}

\begin{table}[th]
\caption{Average AUC (in \%) in 5-fold CV for the classification of \textbf{MCI vs. AD} using BERT and W2V2 features (best layer) from the VFT and PDT in self-attention and cross-attention with and without pause coding (P1--4) and disfluencies (Disfl.).}
  \label{tab:mciad}
  \centering
\begin{tabular}{l||cc|cc}
            \toprule
            & \multicolumn{2}{c|}{\textbf{VFT}}                               & \multicolumn{2}{c}{\textbf{PDT}}                              \\ 
            \midrule
features      & self-att                     & cross-att                       & self-att                     & cross-att                     \\ 
            \midrule
BERT             & 71.0                     & 66.7                        & 71.8                     & 76.4\\
BERT P1          & 68.8                     & 69.2                        & 77.6                     & 76.3                      \\
BERT P2          & 67.5                     & 67.1                        & 77.3                     & \textbf{77.8}\\
BERT P3          & \ul{71.3}                & 68.6                        & 74.7                     & \textbf{77.8}\\
BERT P4          & \textbf{71.7}            & \ul{70.4}                   & \ul{79.5}                & \textbf{77.8}\\
BERT P3 + Disfl. & 70.6                     & \textbf{70.6}               & \textbf{80.5}            & \ul{77.2}\\ 
            \midrule
W2V2             &    67.1                  & -                            & 73.4                     & -                          \\
            \bottomrule
\end{tabular}
\end{table}

\begin{table}[th]
\caption{Average AUC (in \%) in 5-fold CV for the classification of \textbf{NC vs. AD} using BERT and W2V2 features (best layer) from the VFT and PDT in self-attention and cross-attention with and without pause coding (P1--4) and disfluencies (Disfl.).}
  \label{tab:hcad}
  \centering
\begin{tabular}{l||cc|cc}
            \toprule
            & \multicolumn{2}{c|}{\textbf{VFT}}                               & \multicolumn{2}{c}{\textbf{PDT}}                              \\ 
            \midrule
features      & self-att                     & cross-att                       & self-att                     & cross-att                     \\ 
            \midrule
BERT             & 88.3                     & \textbf{86.9}                    & 84.4                     & 84.6                      \\
BERT P1          & \textbf{89.6}            & \textbf{86.9}                    & \textbf{88.2}            & 85.5                       \\
BERT P2          & \ul{88.8}                & 85.5                             & \ul{85.4}                & 84.1                      \\
BERT P3          & 88.0                     & \ul{86.8}                        & 85.2                     & 85.5                      \\
BERT P4          & 87.4                     & 85.7                             & 84.7                     & \ul{86.0}                      \\
BERT P3 + Disfl. & 86.6                     & 85.7                             & 84.9                     & \textbf{86.4}                      \\
            \midrule
W2V2             & 85.3                     & -                                & 83.1                      & -                          \\
            \bottomrule
\end{tabular}
\end{table}

\section{Discussion and Conclusion}
In this work, we investigated the potential of speech pauses as markers of cognitive decline in text-based dementia assessment using speech from two cognitive tests.
We were able to show that our systems are quite capable of classifying cognitive impairment at different stages and benefit from enrichment with pauses.
In addition we investigated the effect of incorporating pause information from the acoustic context into the text-based assessment.
We found that the VFT performed best in detecting MCI versus healthy controls when the text-based assessment could learn from acoustic information.
For the detection of MCI versus dementia, the PDT showed the highest discriminative power, with modeling of pauses and disfluencies being sufficient in the text-based assessment. 
Dementia could be excluded from healthy controls regardless of test and context, but pause modeling is advantageous.

While our findings suggest pauses in speech offer a promising indicator for the detection of cognitive decline in clinical assessment, several key considerations and challenges remain.
%Variability in pause patterns: 
Pause occurrence and duration can be influenced by several factors besides dementia, including age, language, speaking style, and emotional state. 
This necessitates careful control of these biases and potentially incorporating additional speech features to enhance the robustness of analysis.
%Limited generalizability:
Research on speech pause analysis in dementia detection often involves relatively small datasets. 
Further studies with larger and more diverse populations are crucial to establish the generalizability and reliability of this approach for real-world application.

It's crucial to remember that no single biomarker is currently considered a definitive diagnostic tool for dementia.  
Physicians often utilize a combination of approaches, including clinical evaluation, cognitive assessments, and sometimes a combination of these modalities, to reach a diagnosis. 
Additionally, the specific tests chosen and their interpretation depend on various factors, including the patient's individual situation and suspected type of dementia.
Dementia encompasses various subtypes (e.g., Alzheimer's, Lewy body), each potentially showing varying levels of detectability through different modalities and tests.
Combining pause analysis with other speech features (e.g., prosody, articulation) in multimodal and multi-test approaches hold potential for improving the accuracy and robustness of dementia detection.

Overall, the research on speech pauses in dementia detection is promising, offering valuable insights into potential diagnostic markers.
The results also suggest that acoustic cross-attention could benefit from pause-dependent masking to prevent overfitting to other regions.
However, addressing the mentioned challenges and limitations is crucial for establishing this approach as a reliable and generalizable tool for clinical use.

%fine tune language models on large amount of data from different languages and populations
%\section{Acknowledgments}
%To be added after acceptance.

\pagebreak

\bibliographystyle{IEEEtran}
\bibliography{mybib}

\end{document}